**Phonons in ultrathin oxide films – 2D to 3D transition in FeO on Pt(111)**

N. Spiridis,[1] M. Zając,[2,3,a] P. Piekarz,[4] A.I. Chumakov,[2] K. Freindl,[1] J. Goniakowski,[5] A. Kozioł-Rachwał,[3] K. Parliński,[4] M. Ślęzak,[3] T. Ślęzak,[3] U.D. Wdowik,[6] D. Wilgocka-Ślęzak,[1] J. Korecki, [1,3,*]

[1]*Jerzy Haber Institute of Catalysis and Surface Chemistry, Polish Academy of Sciences, ul. Niezapominajek 8, 30-239 Kraków, Poland.*

[2]*European Synchrotron Radiation Facility (ESRF), P.O. Box 220, F-38043 Grenoble, France*

[3]*Faculty of Physics and Applied Computer Science, AGH University of Science and Technology, al. Mickiewicza 30, 30-059 Kraków, Poland*

[4]*Institute of Nuclear Physics, Polish Academy of Sciences, ul. Radzikowskiego 152, 31-342 Kraków, Poland*

[5] *Sorbonne Universités, UPMC Univ Paris 06, CNRS-UMR 7588, Institut des NanoSciences de Paris, F-75005, Paris, France*

[6]*Institute of Technology, Pedagogical University, ul. Podchorążych 2, 30-084 Kraków, Poland*

*korecki@agh.edu.pl

---

[a] present address: National Synchrotron Radiation Centre SOLARIS at Jagiellonian University, ul. Gołębia 24, 31-007 Kraków, Poland

**Abstract**

The structural and magnetic properties of ultrathin FeO(111) films on Pt(111) with thicknesses from 1 to 16 monolayers (ML) were studied using the nuclear inelastic scattering (NIS) of synchrotron radiation. Distinct evolution of vibrational characteristics with thickness that is revealed in the phonon density of states (PDOS) witnesses a textbook transition from 2D to 3D lattice dynamics. For the thinnest films of 1 and 2ML, the low energy part of the PDOS followed a linear $\propto E$ dependence in energy that is characteristic for 2-dimensional systems. This dependence gradually transforms with thickness to the bulk $\propto E^2$ relationship. Density functional theory phonon calculations perfectly reproduced the measured 1ML PDOS within a simple model of a pseudomorphic FeO/Pt(111) interface. The calculations show that the 2D PDOS character is due to a weak coupling of the FeO film to the Pt(111) substrate. The evolution of the vibrational properties with an increasing thickness is closely related to a transient long range magnetic order and stabilization of an unusual structural phase.

Dimensionality has the major impact on the basic properties of the condensed matter. In low-dimensional systems, the confinement of electrons and phonons changes their dispersion relations and density of states, which is reflected in modifications of electrical, magnetic, thermal and optical properties when going from the 3-dimensional (3D) bulk to a 2-dimensional (2D) monolayer.[1] The effect of the reduced dimensionality is relatively simple to observe for the electronic system thanks to the short characteristic screening length and large energy level splitting. Conversely, due to the collective character of lattice vibrations, truly 2D phononic systems are limited to monolayer membranes, like graphene[1] or similar quasi two-dimensional systems.[2] Interest in primary information on the crystal structure that is contained in the vibrational properties rises a demand for direct measurements of the phonon spectra in low dimensional systems. Phonons determine elastic and thermodynamic properties and mediate many coupling phenomena. Phonons also promote phase transitions, which often can be predicted from dispersion relations.[3]

Ultrathin epitaxial oxide films are of recent interest as model catalysts, sensors, in microelectronics and spintronics, and other applications.[4] The extent to which the properties of couple-monolayer films resemble those of their bulk counterparts is an increasingly important question. First-principle calculations show the complexity of the problem and reveal specific features of atomic[5] and electronic structure[6] of ultra-thin films and metal-supported oxide monolayers.[7] Metastable structures can be stabilized by epitaxy as a result of the intricate of different film and substrate parameters.[8] Polarity may be the driving factor of artificial structure stabilization, especially in films composed of a couple of monolayers.[9,10,11] In one of the most extensively studied polar oxide-film systems, i.e. for FeO(111) films on Pt(111),[12] the structural stability could be extended beyond the monolayer range.[13] This gave us opportunity to systematically study the evolution of the vibrational properties characteristic for a 2D to 3D-transition.

We have measured the partial iron phonon density of states (PDOS) as a function of FeO thickness from a single to over a dozen monolayers using the nuclear inelastic scattering (NIS) of synchrotron radiation,[14,15] and we showed that the PDOS is distinctly different from bulk FeO (wüstite). In particular, the monolayer PDOS is characterized by a sharp feature at an energy of E=24 meV, which is well-explained within a density functional theory (DFT) phonon calculations. Furthermore, instead of the Debye-like-$\propto E^2$ behavior, the film PDOS scales linearly with energy in the 1–2 monolayer range, which represents truly 2-dimensional behavior. A gradual transition toward the 3D characteristics with thickness is accompanied by magnetic anomalies that are distinctly reflected in the phonon spectra.

Bulk FeO has an NaCl-type structure, and the (111)-oriented FeO epitaxial films are considered as the fcc stacking of alternating hexagonal iron and oxygen planes.[12] A considerable (10%) lattice mismatch between Pt(111) and FeO(111) (in plane spacing of surface atoms $a_{Pt}$=2.77Å and $a_{FeO}$=3.04Å, respectively) results in a complex misfit moiré structures for 1- and 2-monolayer (ML) FeO films.[16,17,18,19,20,21] The positions of the iron and oxygen atoms are modulated between the fcc and hcp sites of the Pt(111) surface with a periodicity corresponding to the misfit parameter.[16,20] The experimental incommensurate moiré structure has a (9x9) periodicity with respect to Pt(111), additionally indicating a rotational mismatch of several degrees.[16,18] While an unsupported FeO(111) monolayer is perfectly flat, the interaction and charge exchange with the Pt(111) substrate induce a separation of iron and oxygen atomic planes[22] of 0.7Å, which is considerably smaller than 1.24Å in wüstite.[17]

Beyond a thickness of 2ML, the preferential formation of magnetite ($Fe_3O_4$) is observed.[23] However, it was shown recently that the FeO(111) films on Pt(111) can be stabilized to a thickness of several nanometers using a carefully adapted preparation.[13] These films display electronic and magnetic properties markedly different from wüstite. Specifically, the conversion electron Mössbauer spectroscopy (CEMS) indicated a high degree of covalency in the Fe-O bonds and a long-range magnetic order in the thickness range of a few monolayers. It has been suggested that these properties signify a new phase that is stabilized by epitaxial growth. The present study of vibrational properties using the nuclear resonance scattering (NRS) of synchrotron radiation combined with the *ab initio* phonon calculations sheds light on this question.

Experiments were performed in an ultrahigh vacuum system[24] at the ID18-beamline[25] in ESRF Grenoble. $^{57}$FeO films with a thickness *d* in from 1 to 16ML (1ML is equivalent to 2.48Å of bulk FeO) were epitaxially grown on an atomically clean Pt(111) substrate using the preparation protocol described earlier,[13] and the sample structure was verified *in situ* using characteristic LEED patterns. Prior to the NIS experiments the samples were characterized by the coherent grazing-incidence NRS measurements[26]. The $\vec{k}$-vector of the incident X-rays was parallel (within ±3°) to the dense-packed direction on Pt(111). The NRS time spectra shown in Fig.1 together with the best fits using the REFTIM software[27] coincide with the previous CEMS measurements for another set of samples.[13] The quantum beat modulation of the scattered intensity decay that appears at 6ML reveals long-range magnetic order with the average hyperfine magnetic field $B_{hf}$ increasing from $(23.1\pm0.2)$T at 6ML to $(28\pm1)$T at 10ML. Above this thickness $B_{hf}$ dropped to almost zero within few monolayers when high- and low-$B_{hf}$ phases coexisted. For the high-$B_{hf}$ phase, the numerical analysis explicitly indicated the ferromagnetic order with the magnetization along a unique direction, namely at $(30\pm5)°$ with respect to $\vec{k}$. It is not clear whether the transition above 10ML is to the paramagnetic or to the antiferromagnetic state with a low magnetic moment.

The unusual electronic and magnetic properties of the FeO layers are accompanied by atypical NIS spectra (see Supplementary Material 1), which measure the probability of the phonon-assisted nuclear resonance absorption of photons as a function of the detuning energy $\Delta E$ from the 14.412 keV resonance for $^{57}$Fe. The partial iron PDOS, $g(E)$, can be derived from the NIS spectra in a parameter-free procedure, which considers the thermal population of phonons and multi-phonon processes.[28,29] Due to the grazing incidence geometry, PDOS corresponds to phonons with a large in-plane component of the polarization vector along the direction given by the projection of the X-ray $\vec{k}$-vector on the sample surface. The PDOS presented in Fig.2 are distinctly different from that of bulk wüstite,[30,31] The most characteristic features of PDOS are: (i) a sharp peak at 23.5 meV and a sudden cutoff for the thinnest and the thickest films, (ii) a clear correlation between the smearing of PDOS and the magnetic order for intermediate thicknesses, (iii) phonon hardening above the thickness of the magnetic anomaly, and (iv) a deviation from the 3D$\propto E^2$ Debye behavior in the low energy range for the 1 and 2ML films. The power $n$ in the $g(E)\propto E^n$ dependence at low $E$ within the Debye approximation relates to the system dimensionality $D$: $n=D-1$. Hence, $n=2$ is common for 3-dimensional systems, while $n=1$ corresponds to a 2-dimensional behavior. The linear $g(E)$ dependence for low-dimensional systems is the subject of a long debate due to controversies between experiments (NIS or neutron scattering) on different nanoparticle assemblies, for which linear,[32,33,34] quadratic[35,36,37,38] and fractional[39,40,41] energy dependencies have been reported. This diverse behavior reflects the complexity of nanomaterials, for which the effects due to reduced dimensionality, finite size, and different local coordination interplay.[42,43] For a model 2D interfacial system, i.e., for the Fe monolayer on W(110),[44] PDOS, albeit very different in the overall character from bulk Fe, displayed typical $\propto E^2$ Debye behavior in the long wavelength limit. As shown theoretically [34] the prerequisite for the 2-dimensional character of the surface phonons is a decoupling between the surface and the interior, which is not realized for neither the metallic iso-structural systems[44] nor the surface of a metal single crystal.[45]

In contrast, the two-dimensional behavior is demonstrated clearly for our epitaxial FeO films on Pt(111) in the ultrathin limit in Fig.3(top), where the low energy experimental points of $g(E)$ are shown in the double logarithmic plot. In such a plot, the slope of the linear fit represents the power $n$ in the $g(E) \propto E^n$ dependence. The fitted $n=1.0\pm0.05$ unambiguously witnesses of the 2D Debye solid for 1 and 2ML. The power $n$ increases to $n=1.5\pm0.06$ for d=4ML and nearly reaches the 3D limit for d≥9ML, scaling with the reciprocal of the film thickness. This shows that vibrational characteristic of the FeO films is little dependent on the Pt(111) substrate and suggests a weak metal-oxide coupling at the interface. Remarkably, from the point of view of the continuous-medium Debye model, the truly 2D system is not only the monolayer but also the bilayer FeO film, which means that the bulk-like acoustic phonon modes start to develop only above 2ML.

The 2-dimensional characteristic of PDOS for the monolayer FeO films on Pt(111) and the strong deviation from the vibrational bulk-FeO properties are related to the structural specificity of the FeO films, as studied in detail in the monolayer limit both experimentally[16,17,18,46] and theoretically.[20,21] A large misfit between the Pt(111) and FeO(111) planes leads to the moiré structure, which has been explained by the modulation of the position of the iron and oxygen atoms with respect to the platinum substrate . According to Giordano et al.,[20] the strongest FeO-Pt interaction occurs when both the Fe and O atoms are located within the hollow Pt(111) sites and the weakest when the Fe atom is located in the on-top position with respect to a surface Pt atom. For the latter case, the structural properties of the FeO/Pt(111) monolayer are similar to those of a free-standing FeO monolayer.[20]

To elucidate the observed features we compared the measured PDOS with the first principle calculations. The computational model was built of (4x4)-Pt(111) slabs, five atomic layer-thick (80 atoms), covered on one side by the FeO monolayer (32 atoms). The slabs were spaced by a 1.8 nm thick vacuum layer. This structure was fully optimized using the projector-augmented wave method[47] and the generalized-gradient approximation within the VASP program.[48] Strong electron interactions in the 3d states were treated with the DFT+U approach ($U_{Fe}$–$J_{Fe}$ = 3 eV).[49] The supported FeO monolayer is metallic with the dominant contribution of minority-spin 3d states near the Fermi energy (see Supplementary Material 2). We assumed an antiferromagnetic order for the Fe atoms, which provides lower energy than the ferromagnetic and nonmagnetic states.[20] The phonon spectra were calculated using the direct method implemented in the Phonon software.[50] The force constants and dynamical matrices were derived by displacing all non-equivalent atoms from equilibrium positions and calculating the Hellmann-Feynman forces. The PDOS was obtained by randomly sampling the k-points in the first Brillouin zone.

As to account for the large unit cell of the actual FeO/Pt(111) interface, we adopted a strategy proposed by Giordano et al.[20] Three different pseudomorphic configurations corresponding to the different regions of the moiré unit cell were considered.. As a result of optimization , the only stable pseudomorphic structure (without soft phonon modes), was the "Fe-fcc" (both O and Fe ions in three-fold hollow sites of a Pt(111) substrate). In the course of optimization, the "Fe-hcp" (O ions on-top of surface Pt atoms, Fe ions in the hollow sites) structure relaxed to "Fe-fcc," and the "Fe-top" (O ions in the hollow sites, Fe ions on-top of a surface Pt) structure relaxed to a new atomic configuration, in which Fe atoms occupied intermediate positions. It can be concluded that only the fcc regions stabilize the moiré structures and we will focus on the phonon spectra obtained for this region in the following.

The calculated partial Fe PDOS functions for the "Fe-fcc" structure decomposed in the X, Y (in-plane) and Z (normal) directions are shown in Fig. 4 (PDOS for the "Fe-top" region is presented in Supplementary Material 3). The vibrational characteristics are strongly anisotropic, and the Y component can be directly compared with the experimental results in Fig.4b because the $\vec{k}$ vector of the X-rays in the NIS experiment was nearly parallel to the Y-direction. Despite the simplifications in the theoretical model, the agreement between the experiment and theory is extremely good. This suggests that the substrate-induced modulation of the film structure within the moiré pattern influences only little the vibrational properties of the FeO layer and confirms a weak metal-oxide coupling at the interface. Additionally, the calculations reveal high energy vibration of the Fe atoms above 47 meV that were not observed in the bulk rock-salt structure.[31]

The explanation of the vibrational properties of the thicker FeO layers is less straightforward, because their increased structural complexity is a severe limitation to theoretical modeling. However, aside from the magnetic anomaly that is discussed below, similarity of PDOS over the whole thickness range from 1ML to 16ML is obvious, which implies structural affinity of the thicker films to the monolayer. The most characteristic features: the peak at approximately 25 meV and the sharp cutoff above this energy are preserved for the thickest (4 nm) FeO film, with only a slight phonon stiffening as compared to the monolayer limit. Such a blue shift can be generally expected as a result of the enhanced system stiffness, however, it was recently shown that the detailed balance of different factors contributing to the force constants in oxide systems is complex.[51] Nevertheless, the distinct dissimilarity to the phonon spectra of bulk wüstite[30] or to epitaxial magnetite films[52] clearly indicates that a new artificial FeO structure was stabilized on Pt(111). The special character of the vibrational spectra in the ultrathin oxide films may have strong implications for all processes involving phonons (thermodynamic and transport properties), and is of the crucial importance for the chemical reactivity.[53]

Finally, we discuss the magnetic anomaly reflected in the phonon spectra for 6 to 10ML. A magnetic transition, whether or not accompanied by a structural one, can substantially modify the vibrational properties in the bulk, by changing all phonon characteristics; frequency, intensity, and spectral line width.[54] The long range magnetic order may be also responsible for the stabilization of the specific crystal structure,[55] which is of particular importance for epitaxial metastable systems.[56] For ultrathin fcc-Fe(001) films on Cu(1001) Benedek et.al.[57] showed a sharp increase of the phonon frequency below the Curie temperature. For Fe(110) films on W(110)[44] the influence of the ferromagnetic order on the phonon spectra was masked by strong structural effects. For the present case a model situation takes place, in which the magnetic order is established with only minor structural changes, as judged from the LEED pattern. Hence, we interpret the broadening in the phonon spectra for 6 to 10ML as the effect of the spin ordering, which may result in magnetostrictive strains and a lattice distortion. When the ferromagnetic order vanishes above 10ML, the PDOS sharpens again, however the main peak remains by a factor of two broader than the exceptionally narrow peak for 1ML, which has the full width at half maximum approaching the experimental resolution of 2 meV. Similar situation was observed also for MnO(100) films on Pt(111).[58] The thickness-dependent PDOS broadening results from the increased structural complexity of the thicker FeO films due to a possible modulation of the interplanar distances. The detailed determination of the thick-FeO film structure, very different from its bulk counterpart – wüstite, is the subject of the ongoing research.

**Acknowledgments**

This work was supported by the National Science Center Poland (NCN) under Project No. 2011/02/A/ST3/00150). PP acknowledges support by NCN under Project No. 2012/04/A/ST3/00331 and COST Action MP1308 "Towards Oxide-Based Electronics". The research was performed in the framework of the Marian Smoluchowski Krakow Research Consortium- Leading National Research Centre (KNOW), which is supported by the Ministry of Science and Higher Education of Poland.

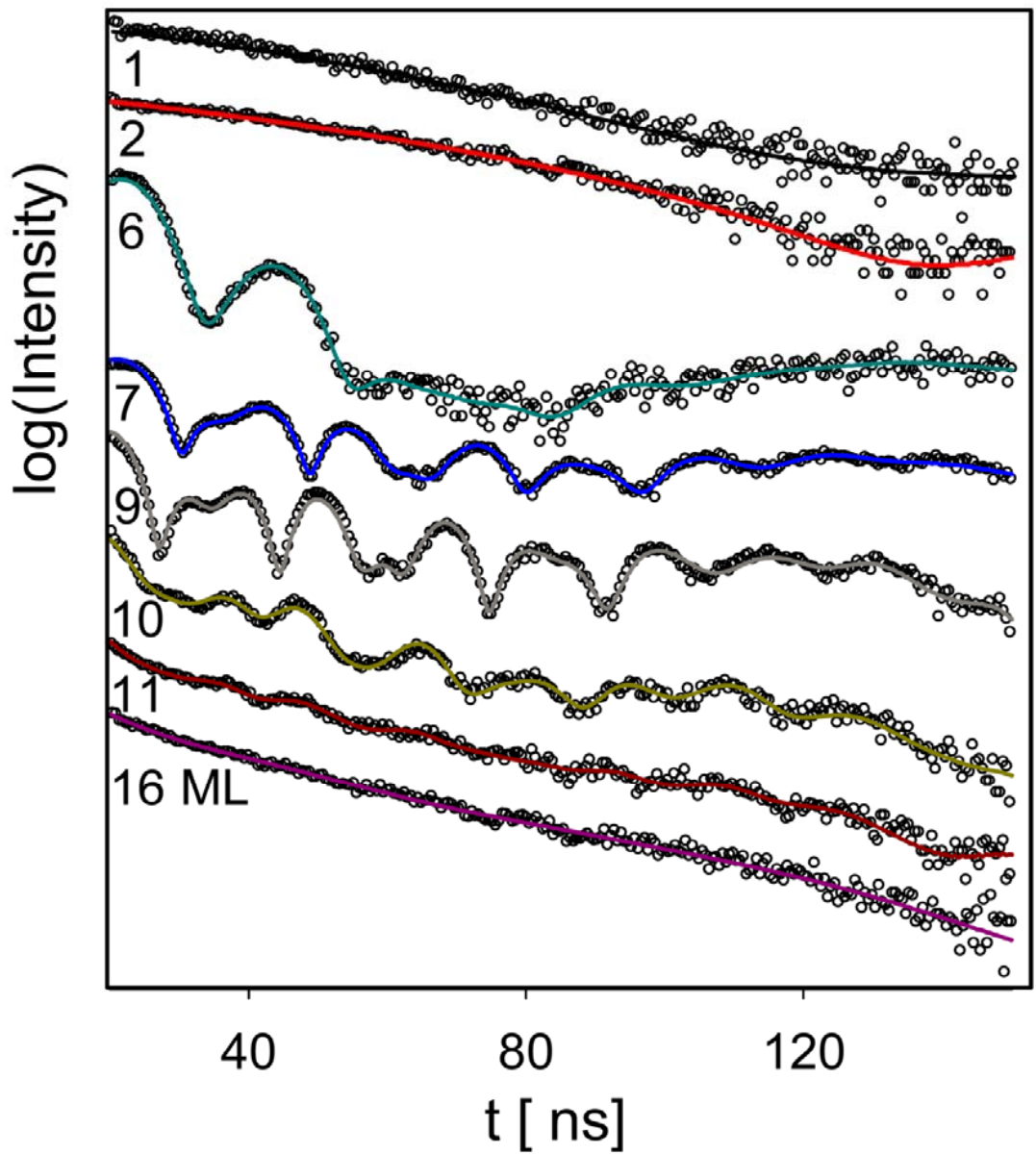

Fig.1 (color online). NRS time spectra for $^{57}$FeO/Pt(111) as a function of FeO thickness.

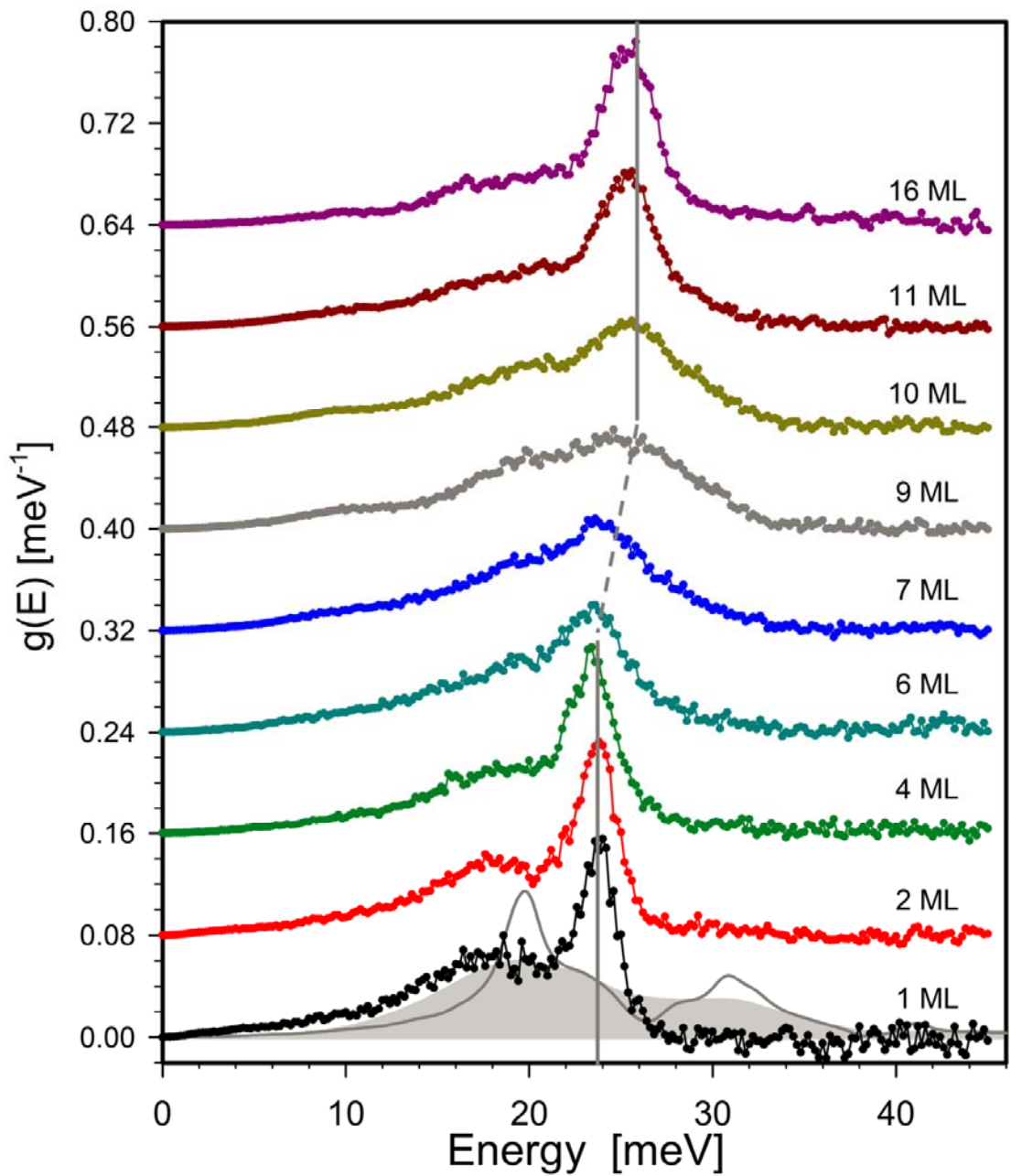

Fig.2 (color online). Phonon DOS *g(E)* for FeO(111) films on Pt(111) as a function of thickness. For clarity, the plots are vertically shifted by 0.08 meV$^{-1}$. The vertical lines mark the shift of the high energy phonon peak. PDOS for 1ML is compared with theoretical calculations[31] for stoichiometric (solid line) and non-stoichiometric wüstite.

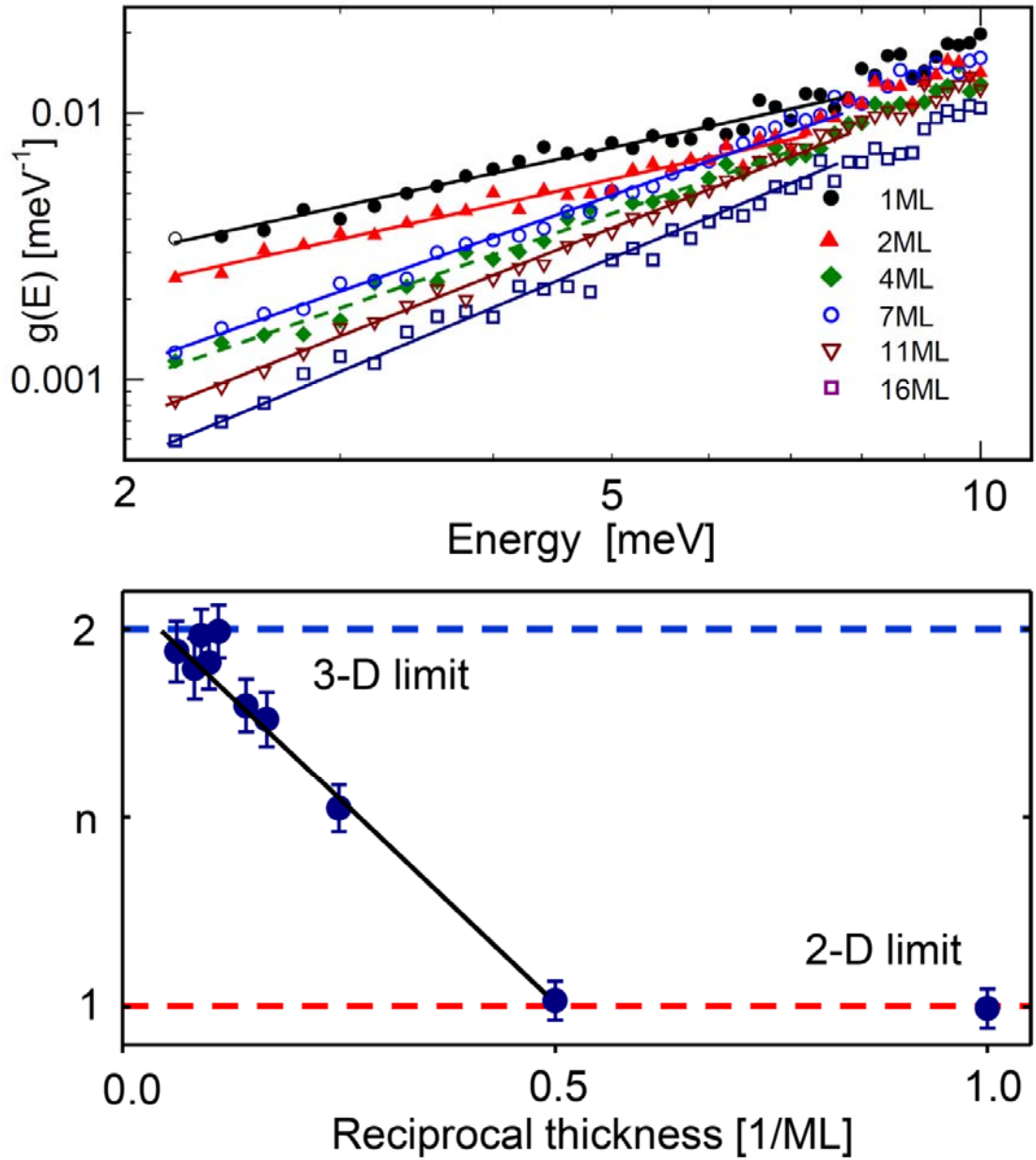

Fig.3 (color online). (Top) Double logarithmic plot of PDOS $g(E)$ for low energy. The straight lines are the fits of the experimental points to a $g(E) \propto E^n$ dependence. (Bottom) The plot of $n$ as a function of the reciprocal of the FeO film thickness $d$ (in ML). The lines at $n=1$ and $n=2$ are the 2D and 3D limits, respectively.

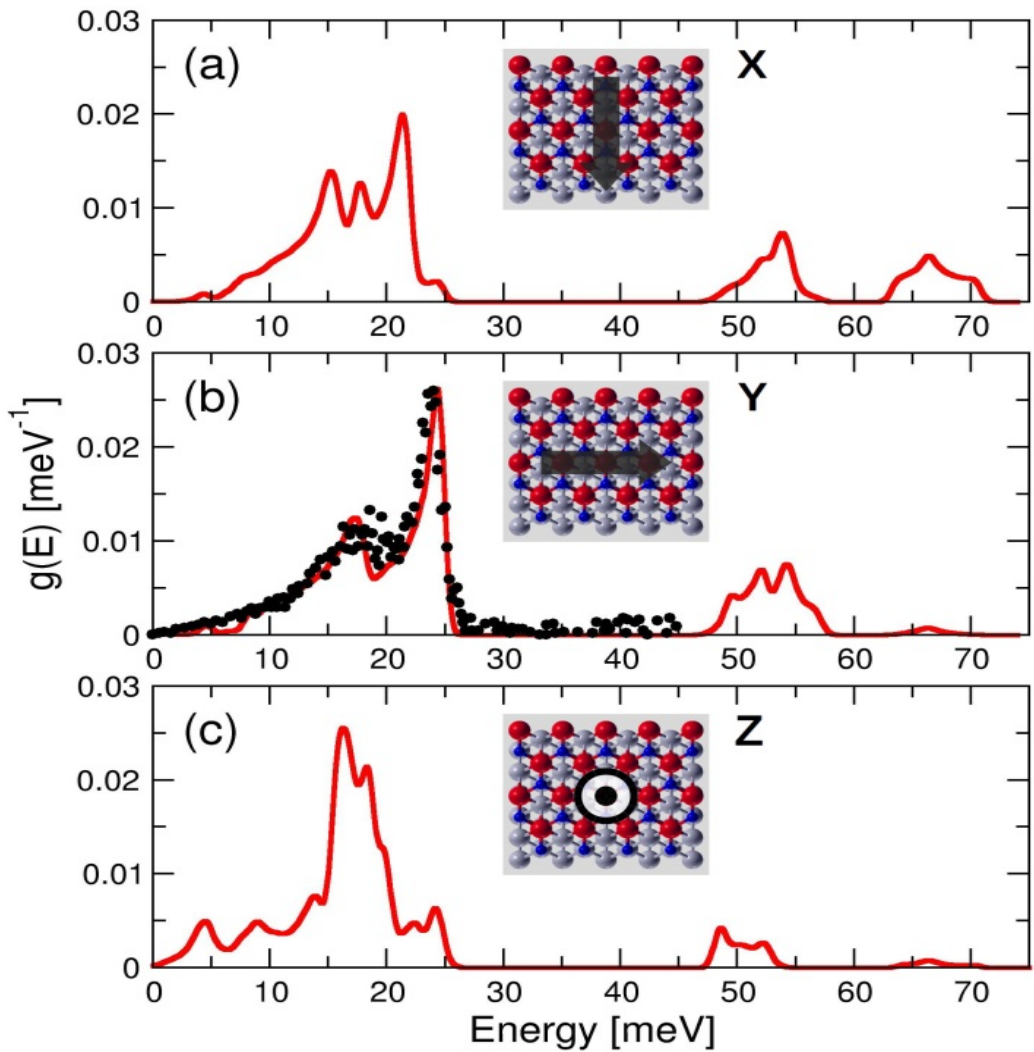

Fig.4 (color online). Partial Fe PDOS for the "Fe-fcc" structure of the FeO monolayer decomposed in the X, Y and Z directions. Insets depict the X-Y-Z orientation on a top view of the stick-and-ball model of the 1ML FeO/Pt(111) system (Pt - light gray, Fe - red/large dark gray, O - blue/small dark gray). The solid lines show the results of DFT calculations, the points in (b) show the experimental data.

**Supplementary Material 1**

An example of the measured spectra for the thickest 16 ML FeO film is shown in Fig. S1 (top) on a logarithmic scale (black curve) to visualize the elastic peak, which is more intense by two orders of magnitude than the phonon-assisted events.

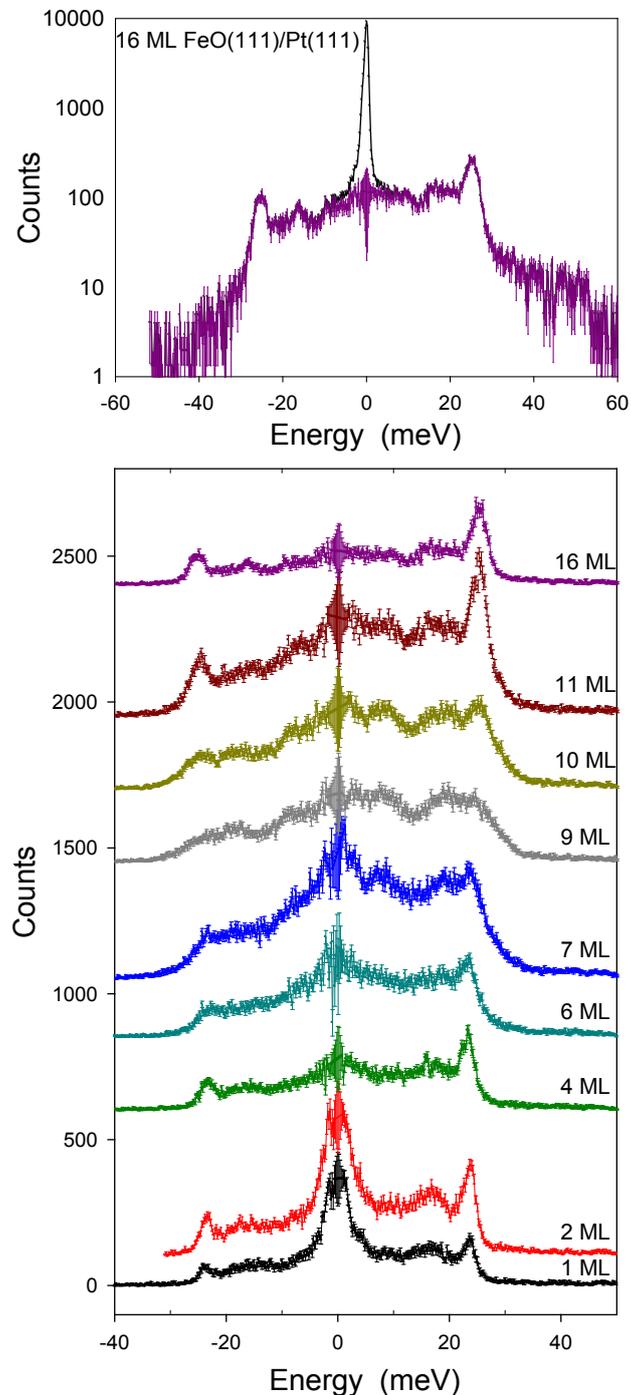

Fig. S1. (Top) The NIS spectrum for the 16 ML FeO film (continuous line), shown on a logarithmic scale to visualize the elastic peak. The error bars represent the statistical errors and the errors due to the uncertainty of the elastic peak subtraction procedure. (Bottom) NIS spectra after subtraction of the elastic peak as a function of the FeO film thickness. The spectra are vertically offset for clarity.

The elastic peak must be subtracted for further analysis, as shown in Fig. S1 (bottom), where error bars have been added to the experimental points. Large error values near ΔE=0 are the result of the elastic peak subtraction, and the elastic peak range, i.e., the energy range of approximately ±2 meV, corresponding to the instrumental energy resolution, should be excluded from the analysis. The NIS spectra in Fig. S1 (bottom) show clear changes with the FeO thickness. The most characteristic is the strong intensity enhancement at low energies for the thinnest films (1 ML and 2 ML), and the sharp peak at approximately 25 meV, which got smeared in the thickness range where the magnetic order is observed.

**Supplementary Material 2**

Figure S2 presents the electron density of states for the "fcc" FeO/Pt(111) monolayer decomposed into three main contributions: Fe-3d, O-2p, and Pt-5d states. The upper and lower pannels show the majority-spin and minority-spin states, respectively, in the antiferromagnetic ground state (given per one atom). The Pt-5d states are shown only for the top surface layer. The FeO monolayer is metallic with the minority-spin states dominating at the Fermi energy and the main majority-spin band shifted about 3 eV to lower energies due to the Hubbard U interaction. The main minority-spin band is located about 1 eV above the Fermi energy.

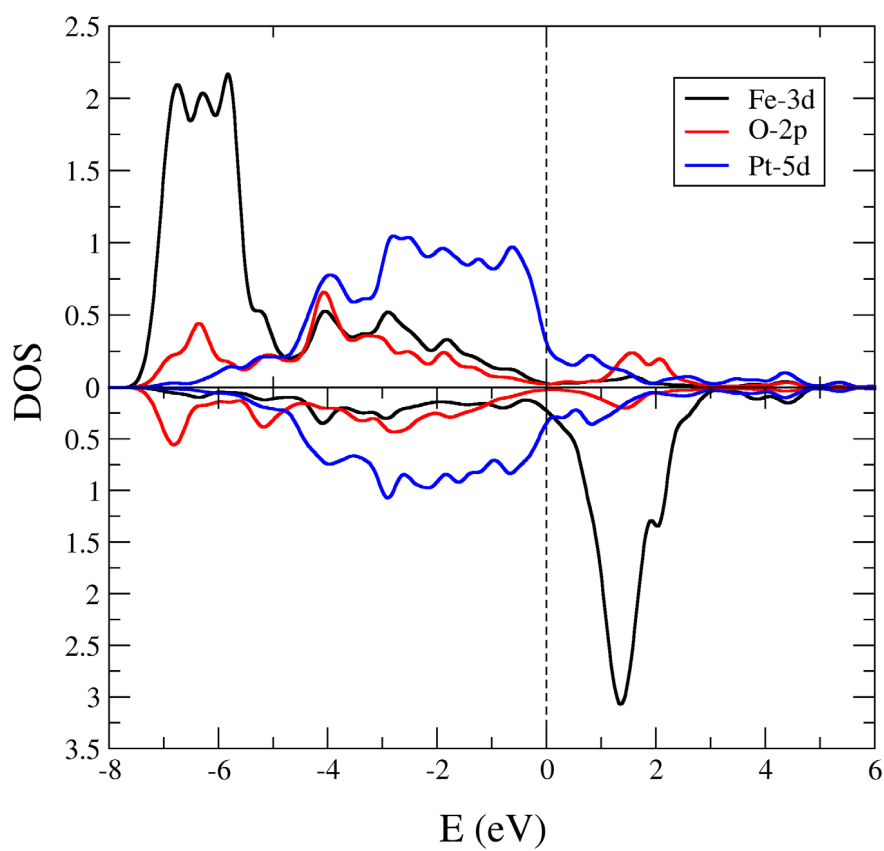

Fig. S2. Spin polarized electron density of states in the "fcc" FeO/Pt(111) layer.

**Supplementary Material 3**

The calculations have been performed also for the "top" pseudomorphic FeO layer, which resulted in a shift of the Fe atoms from ideal positions above the Pt atoms. The calculated partial iron phonon density of states presented in Fig. S3 reveals the modes with imaginary energies (soft modes), which correspond to a rigid shift of FeO plane with respect to the Pt surface. Such modes indicate a weaker coupling between the FeO layer and the surface within this geometry comparing to the "fcc" symmetry. The in-plane Fe vibrations show similar spectra as in the "fcc" structure with a small shift to lower energies induced by the soft modes.

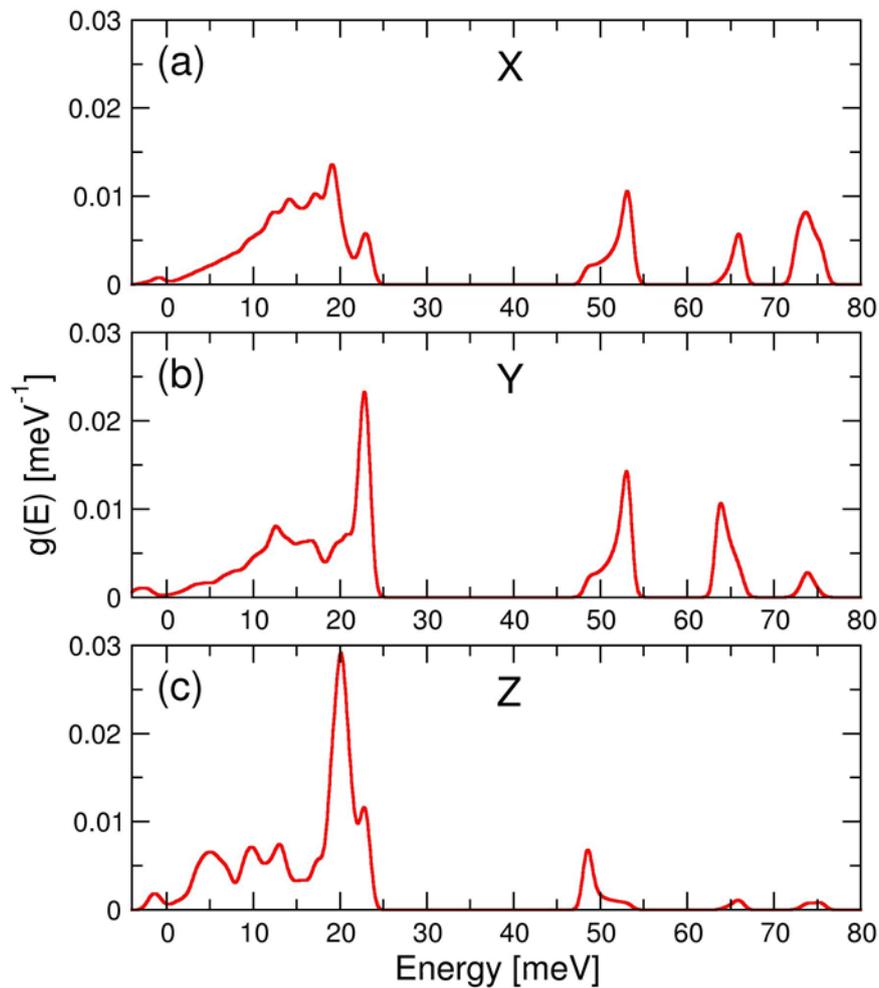

Fig. S3. Phonon partial density of states for the FeO/Pt(111) monolayer within the "top" geometry. The in-plane Fe vibrations along X and Y directions are shown in (a) and (b), respectively, while Z component in (c).